\documentclass[10pt,journal,compsoc]{IEEEtran}

\usepackage{amsmath,amsfonts}
\usepackage{algorithmic}
\usepackage{algorithm}
\usepackage{array}
\usepackage{hyperref}
\usepackage{xurl}
\usepackage[caption=false,font=normalsize,labelfont=sf,textfont=sf]{subfig}
\usepackage{textcomp}
\usepackage{stfloats}
\usepackage{url}
\usepackage{verbatim}
\usepackage{graphicx}
\usepackage{cite}
\usepackage{multirow}

\title{\LARGE \bf
Designing an Optimized Electric Vehicle Charging Station Infrastructure for Urban Area: A Case study from Indonesia
}

\author{Nissa Amilia$^{1}$, Zulkifli Palinrungi$^{2}$, Iwan Vanany$^{3}$, Mansur Arief$^{4}$%
\thanks{$^{1}$Nissa Amilia is with the Department of Technology Management, Institut Teknologi Sepuluh Nopember (ITS), Surabaya, Indonesia}%
\thanks{$^{2}$Zulkifli Palinrungi is with the Department of Electrical and Computer Engineering, Carnegie Mellon University, Pittsburgh, PA, USA}%
\thanks{$^{3}$ Iwan Vanany is with the Department of Industrial and Systems Engineering, Institut Teknologi Sepuluh Nopember (ITS), Surabaya, Indonesia}%
\thanks{$^{4}$Mansur Arief are with the Department of Mechanical Engineering, Carnegie Mellon University, Pittsburgh, PA, USA}%
}

\begin{document}

\maketitle
\thispagestyle{empty}
\pagestyle{empty}

\begin{abstract}

    The rapid development of electric vehicle (EV) technologies promises cleaner air and more efficient transportation systems, especially for polluted and congested urban areas. To capitalize on this potential, the Indonesian government has appointed PLN, its largest state-owned electricity provider, to accelerate the preparation of Indonesia’s EV infrastructure. With a mission of providing reliable, accessible, and cost-effective EV charging station infrastructure throughout the country, the company is prototyping a location-optimized model to simulate how well its infrastructure design reaches customers, fulfills demands, and generates revenue. In this work, we study how PLN could maximize profit by optimally placing EV charging stations in urban areas by adopting a maximal covering location model. In our experiments, we use data from Surabaya, Indonesia, and consider the two main transportation modes for the locals to charge: electric motorcycles and electric cars. Numerical experiments with 11 candidate EV charging station locations and the projected number of electric vehicles in the early penetration phase across 98 sub-districts throughout the city show that only four charging stations are needed to cover the whole city, given the charging technology that PLN has acquired. However, consumers' time-to-travel is exceptionally high (about 35 minutes), which could lead to poor consumer service and hindrance toward EV technologies. Sensitivity analysis reveals that building more charging stations could reduce the time but comes with higher costs due to extra facility installations. Adding layers of redundancy to buffer against outages or other disruptions also incurs higher costs but could be an appealing option to design a more reliable and thriving EV infrastructure. The model can provide insights to decision-makers to devise the most reliable and cost-effective infrastructure designs to support the deployment of electric vehicles and much more advanced intelligent transportation systems in the near future.

\end{abstract}

\section{Introduction}

The ever-increasing use of fossil fuels in Indonesia, the largest economy in Southeast Asia, is one of the main contributing factors to the poor air quality problems numerous cities face. Energy consumption for the transportation sector is estimated to double in the coming years, which is alarmingly high despite government efforts to promote green energy and energy conservation   \cite{energytransindo}. The recent increase in gasoline prices globally exacerbates the problem \cite{gasprice}, requiring the government to provide more than 110 trillion Rupiahs in energy subsidy in the coming years \cite{energysubsidy}, putting severe economic and environmental challenges to government and the public.

The rapid advances in electric vehicles (EVs) and their
advantages over fossil-fueled vehicles promise potential solutions to alleviate the challenges due to fossil fuels and achieve more energy-efficient and environmentally-friendly transportation systems \cite{yong2015review}. With its fast-growing trend globally, rising tenfold from less than half-million units in 2014 to about 4.79 million units in 2019        \cite{evoutlook}, EV has become an appealing technology to adopt at scale. The Indonesian government aims to build reliable facilities and nurture ecosystems for EVs to thrive in the country \cite{maghfiroh2021current}. To carry out the mission, the government appoints  PT PLN Persero (or PLN, in short) to optimally determine the number and location of EV charging facilities in the country. The main expectation is that the charging stations are placed optimally throughout urban areas to maximize demand fulfillment and public accessibility to attract early adopters and smoothen the transition to 100\% electrified urban transportation systems.

Traditionally, facilities and infrastructure planning in Indonesia mainly consider the crowd and intensity level of business and economic activities in an area. This results in facilities concentrating only on business centers or blocks of government buildings, lowering the level of service for these facilities. The public often bears the burden of having to travel long distances to utilize these facilities, which is counterproductive to reducing energy consumption in the first place. 

\begin{figure}[t]
    \centering
    \includegraphics[width=\linewidth]{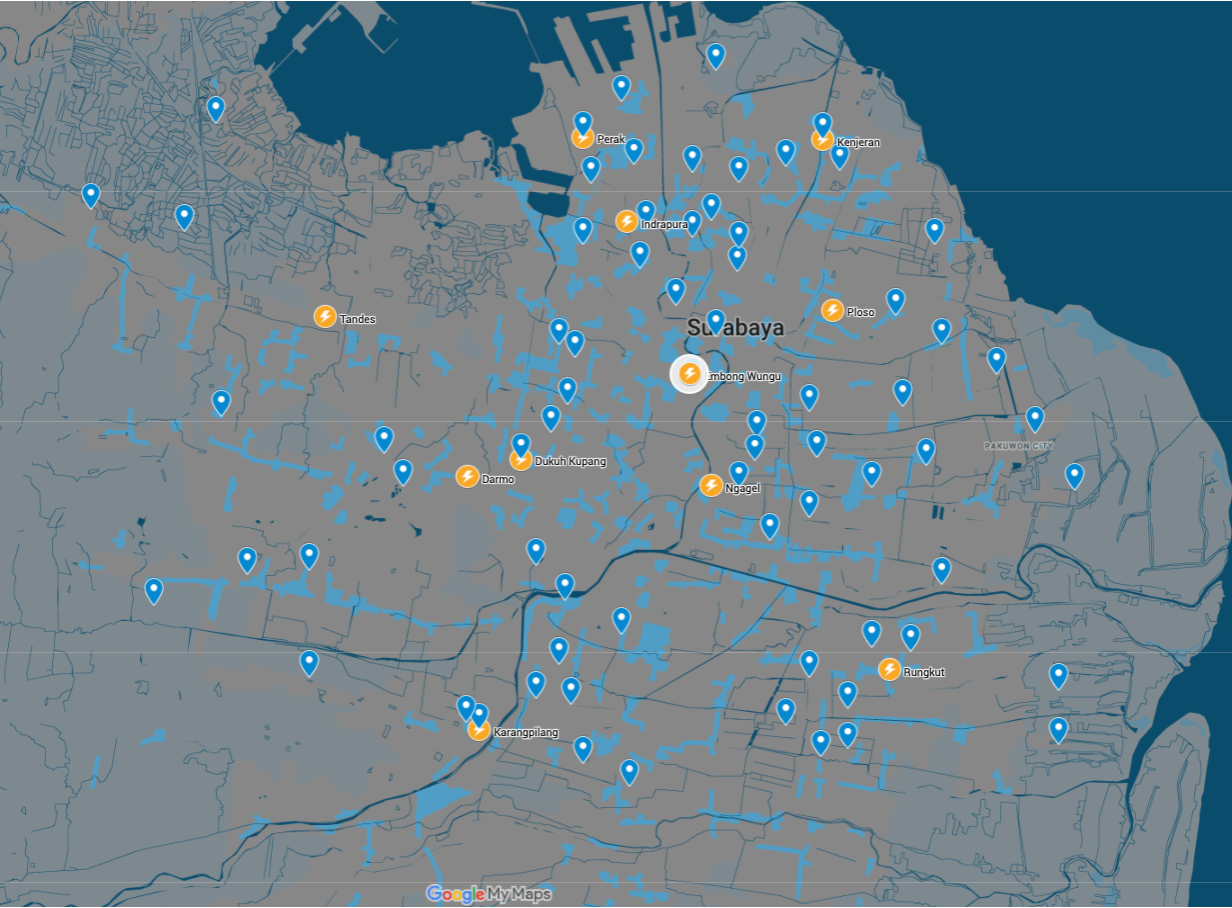}
    \caption{The candidate locations of charging station facilities (yellow marker) and demand points (blue marker) in the city of Surabaya, Indonesia}
    \label{fig:candidate_location}
\end{figure}

In this work, we develop a mixed-integer programming (MIP) model to optimize the location of EV charging stations in urban areas. We collect data from the city of Surabaya and consider 11 alternative locations to install EV charging facilities. The 11 alternative locations are pre-chosen because these represent the sites where PLN can serve customers directly and already have a physical infrastructure deemed feasible to build EV charging facilities (see Fig. \ref{fig:candidate_location} to see how these points distribute across the city). We consider two private transportation modes prevalent in Surabaya: electric cars and electric motorcycles, each with a different charging capacity. In our experiment, we use the projection of each mode as demands in each of the 98 sub-districts in Surabaya and succeed in obtaining the optimal location of the EV charging stations with maximized demand fulfillment for various distance-to-travel constraints. Decision-makers can use the insights to develop a consumer-friendly infrastructure design with maximal revenue to support EV ecosystems on a larger scale.

The contribution of this work can be summarized as follows. First, we present a study case on EV facilities design for an urban area from a developing country's perspective, a potentially colossal EV market shortly. More concretely, we consider reliability issues such as uncertain outages or blackouts in cities in developing countries and add layers of redundancy to buffer against them. In the experiment, we contrast the solutions obtained by not requiring extra coverage from other facilities (Fig. \ref{fig:sumary_opt}) with one that requires such redundancy (Fig. \ref{fig:sumary_opt_2} and Fig. \ref{fig:sumary_opt_3}). Second, we propose a quantitative model based on the maximal covering location model to obtain the optimal facility locations for EV charging stations in urban areas. Finally, we analyze the sensitivity of the optimal locations for an EV charging station concerning customer distance-to-travel and highlight the tradeoff between cost or revenue objective and level of service. These contributions add to the literature optimization model and insights to design effective EV infrastructure for urban areas and thriving EV ecosystems on a larger scale.

The rest of this paper is organized as follows. In Section \ref{sec:problem_formulation}, we present the problem formulation and provide a short overview of related work. We explain our framework and model in Section \ref{sec:framework} and describe our experiment in Section \ref{sec:exp}.  Finally, we discuss our findings in Section \ref{sec:discussion} and conclude in Section \ref{sec:conclusion}.

\section{Related Work}
\label{sec:problem_formulation}

This study deals with facility locations for EV fast-charging infrastructure for urban areas from a developing country perspective. Below we highlight a few selected related works that inspire our formulation. 

Much work on the EV facility location problem considers the new facilities to be integrated into a smart-grid design \cite{chen2012iems, lam2012multi, lam2012capacity} or other renewable energy sources, such as solar cell \cite{guo2012study}. While this view provides an integrated solution to the renewable energy issues to amplify the positive impact of EVs on the environment, this setting is too ideal for urban areas in developing countries.

Case studies on developed and developing countries can be found in EV facility design literature. Research in \cite{frade2011optimal}, set in Lisbon, addresses a similar problem to ours but focuses on slow-charging technology, motivated by the fact that vehicles around the city are often parked overnight. The model proposed in \cite{huang2016design} considers both fast- and slow-charging technologies and uses data from Toronto, focusing on robustly covering all demands, avoiding leaving some demands to be fulfilled only partially. A study case in Ankara uses a GIS-based model and adopts a fuzzy approach  \cite{erbacs2018optimal}. A city-scale simulation is developed in \cite{bi2017simulation} using data from Singapore, focusing on the tradeoff between cost minimization and customer accessibility maximization objective.

Research related to optimizing the location of EV charging station that considers characteristics of urban areas and incorporate the uncertainty and outage of electricity, a common phenomenon in developing countries, is still limited. The earlier study develops a min-max facility location problem to optimize the number of public gas stations in the West Surabaya area that considers the population and traffic densities and number of public facilities (such as hospitals and schools) \cite{rohmat2016penggunaan}. A mixed-integer linear model is utilized to minimize the installation cost of building charging stations, considering the number of vehicles in the area \cite{arayici2019optimal}. An agent-based simulation model is developed in \cite{shi2014study} that considers the number of EVs visiting to charge every hour at each installed charging station, allowing more dynamic analysis but requiring extensive effort to build and validate the model.

Furthermore, work in \cite{adacher2018heuristics} uses $p$-median facility location model for two different objective functions. The model aims to maximize the profit of charging stations under fixed cost while maximizing consumer travel satisfaction, represented by a goal of minimizing the maximum distance between users and charging stations. 
Another work in \cite{ravlic2016optimizing} uses Multi-Source Weber Problem to minimize the total distance from users to their nearest charging stations.
Work in \cite{davidov2016optimization} uses a set covering formulation and discrete location theory to minimize the total cost, considering variable driving behavior (traffic flow), driving range, charging facility installation cost, and road network. 
Finally, \cite{xia2017location} uses a quantum particle swarm optimization algorithm to minimize the cost of using electric vehicles, considering two cost components comprising user costs (including the cost of charging electric vehicles and costs due to travel and waiting for charging) and the charging station costs (including infrastructure construction costs,  (land acquisition costs, equipment, and electrical component cost, and charging station management costs). A good EV charging station facility location review is available in \cite{zhang2019review}.

In this study, we adopt a location set covering problem, which aims at identifying the minimum number and location of facilities to meet customers' demands best \cite{daskin2011network}. This model is suitable for problems aiming to determine the number and assignment of facilities to meet demand points. One fundamental model property is the pre-set coverage radius parameter (often expressed as distance or time threshold), which determines the feasibility of assigning a specific demand node to facilities. This particular threshold value is used to evaluate how large the population or demands can be covered and reached by each facility. It is often helpful to pre-set the facility level of service (the farther the customers are, the more likely the service to be lower since the customers have to wait or travel longer to be serviced by the facility).

\section{Formulation and Proposed Model}
\label{sec:framework}

We consider a set of demand points $I$ and supply stations $J$, representing sub-district regions and charging station candidate locations in an urban area. We also consider $K$ vehicle types, incorporating types of vehicle modalities that urban cities accommodate  (here, we include electric motorcycles and cars). The average time to travel from demand $i \in I$ to charging station $j \in J$ is denoted by $d_{ij}$. A threshold parameter $d_{max}$ is used to limit this time in the following analysis to see the robustness of the solution w.r.t. consumer time-to-travel for charging. The decision variables associated with these points are binary variables
\begin{equation}
x_j = \begin{cases}
1, & \text{if station } j \text{ is selected} \\
0, & \text{otherwise}
\end{cases}
\end{equation}
and 
\begin{equation}
y_{ij} = \begin{cases}
1, & \text{if EVs from } i \text{ are served by station }  j\\
0, & \text{otherwise},
\end{cases}
\end{equation}
indicating whether a charging station candidate location $j$ is selected or not and whether demand point $i$ is to be fulfilled by charging station $j$, respectively. In addition, we also use integer decision variables $v_{ij}^k$ and $u_j$, denoting the number of electric vehicles of type $k$ from point $i$ charged at station $j$ and the number of units of charging connectors installed at charging station $j$, respectively.

Each opened station $j$ incurs a daily cost $h_j$ and can only accommodate $q_j$ charging connectors due to limited space. Each charging connector incurs $g$ daily cost and has a limited daily charging throughput of $c_j$ kWh. For a vehicle type $k$, it takes $e_k$ kWh energy and $t_k$ time to charge using fast-charging technology that PLN adopts fully. To convert the energy used to monetary value, we use the Indonesian electricity price denoted by $r$ Rupiah/kWh (Rupiah or Rp. is Indonesian currency). 

With this setting, the objective is to maximize the daily profit
\begin{equation}
    \text{maximize }  \underbrace{\sum_{i \in I} \sum_{j \in J} \sum_{k \in K} r e_k v_{ij}^k}_{\text{revenue}} - \left( \underbrace{g \sum_{j \in J} u_j + \sum_{j \in J} h_j x_j}_{\text{cost}}\right), \label{obj}
\end{equation}
which takes into account daily revenue and operational and investment costs that have been broken down to daily nominal costs (assuming five years depreciation schedule). This objective is maximized subject to the following set of constraints:
\begin{align}
    \sum_{k \in k} v_{ij}^k &\leq y_{ij} M, &\forall i \in I, j \in J, \label{numberassigned}\\
    d_{ij} y_{ij} & \leq d_{max} , &\forall i \in I, j \in J, \label{distmax}\\
    \sum_{j \in J} v_{ij}^k & = w_i^k, & \forall i \in I, k \in K,
    \label{demandfulfilled}\\
    \sum_{i \in I} \sum_{k \in K} t_k v_{ij}^k  &\leq c_j u_j, & \forall j \in J, \label{charging_cap}\\
    u_j &\leq  x_j q_j, & \forall j \in J, \label{capacityopened}\\
    \sum_{i \in I} y_{ij} & \leq x_j M, & \forall j \in J, \label{assignedopened}\\
    \sum_{j \in J} y_{ij} & \geq 1, & \forall i \in I, \label{demandassigned}\\
    \sum_{j \in J} x_j & \leq N, \label{maxstations}\\
    x_1 &= 1 \label{oneopen}.&
\end{align}

In the above formulation, constraint (\ref{numberassigned}) ensures that charging stations can only charge vehicles if assigned. Constraint (\ref{distmax}) ensures the maximum time-to-charge for consumers does not exceed the set threshold $d_{max}$. Constraint (\ref{demandfulfilled}) ensures all charging demands are fulfilled, where $w_i^k$ denotes the number of vehicles of type $k$ to charge at demand point $i$. Constraint (\ref{charging_cap}) ensures that the required charging capacity to fulfill each station's assigned demand does not exceed the installed capacity. Constraint (\ref{capacityopened}) restricts the number of charging connectors installed in each station. Constraint (\ref{assignedopened}) ensures that demands are assigned only to opened stations. Constraint (\ref{demandassigned}) guarantees that at least one stations cover each demand. Constraint (\ref{maxstations}) limits the maximum number of stations to open. Finally, constraint (\ref{oneopen}) enforces that Station 1 (which is the main EV charging station in Surabaya operated by PLN) open (as demanded by PLN).

In addition, we also have a few variable type constraints
\begin{align}
    x_j &\in \{0, 1\}, &\forall j \in J, \label{xbin}\\
    y_{ij} &\in \{0, 1\}, &\forall i \in I, j \in J, \label{ybin}\\
    v_{ij}^k & \in \{0, 1, 2, \cdots\}, &\forall i \in I, j \in J, k \in K, \label{vnat}\\
    u_j & \in \{0, 1, 2, \cdots\}, &\forall j \in J. \label{unat}
\end{align}
This formulation uses linear objective function and linearized constraints, which yields a mixed-integer programming (MIP) model, allowing us to solve it efficiently using standard MIP solvers.

\section{Numerical experiments}
\label{sec:exp}

We run the model presented in Section \ref{sec:framework} using data collected from Surabaya and interviews with PLN and EV stakeholders in the city. We combined vehicle registration data with the projection of the number of electric motorcycles ($k=1$) and electric cars ($k=2$) for the early penetration phase in Surabaya, distributed on each of its 98 sub-districts as our demand points $u_i^k$'s. We accumulate the number of EVs on a sub-district level and use sub-district coordinates on Google Maps as our demand points to remove personally identifiable information and maintain confidentiality. Furthermore, we prepopulate 11 candidate locations of EV charging stations and enforce Station 1 to open, based on PLN inputs, reflecting the current conditions in the field. 

We also obtain the following information. The estimated daily cost to open a charging station $h_j =$ Rp. 403,288 $\forall j \in J$, and the estimated daily cost to install a charging connector $g =$ Rp. 110,244. PLN fast-charging technology takes 20 minutes to fully charge an electric motorcycle and about 90 minutes for an electric car. At the time of writing, the current electricity rate for business uses is $r=$ Rp. 2,644.78/kWh. We use information from Google Maps to estimate the average travel time $d_{ij}$ from demand $i$ to station $j$. Finally, we set $N=11$ for the maximum number of charging stations, $q_j = 10, \forall j \in J$ for the maximum charging connectors, and $M=1000$ as a practical value for our big-$M$ constraints.

\begin{figure}[t]
    \centering
    \includegraphics[width=\linewidth]{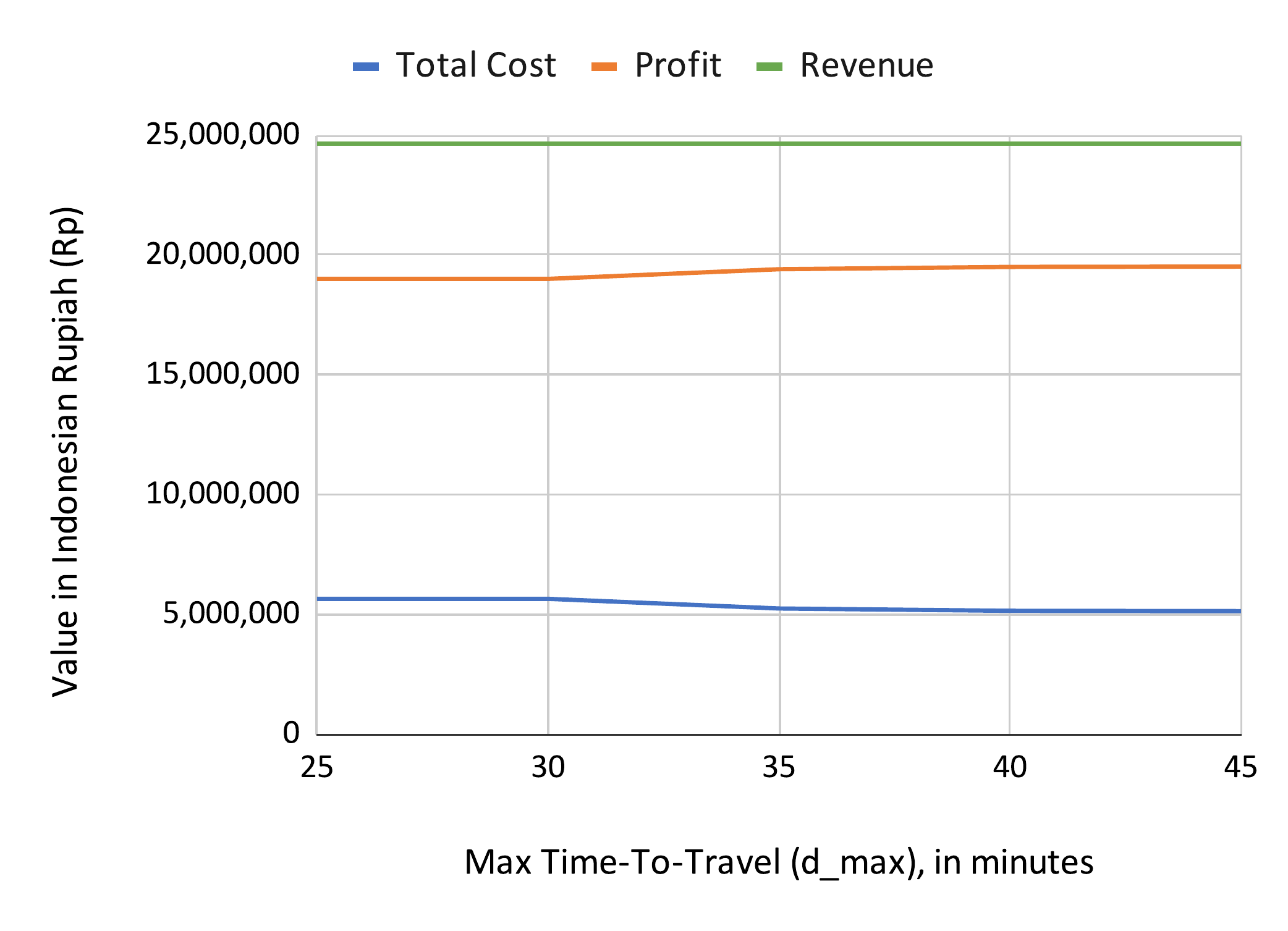}
    \caption{Daily revenue, cost, and profit from the optimal solution of EV charging stations in Surabaya for $d_{max} = \{25, 30, 35, 40, 45\}$ minutes for baseline model (without redundancy)}
    \label{fig:sumary_opt}
\end{figure}

\begin{figure}
    \centering
    \includegraphics[width=\linewidth]{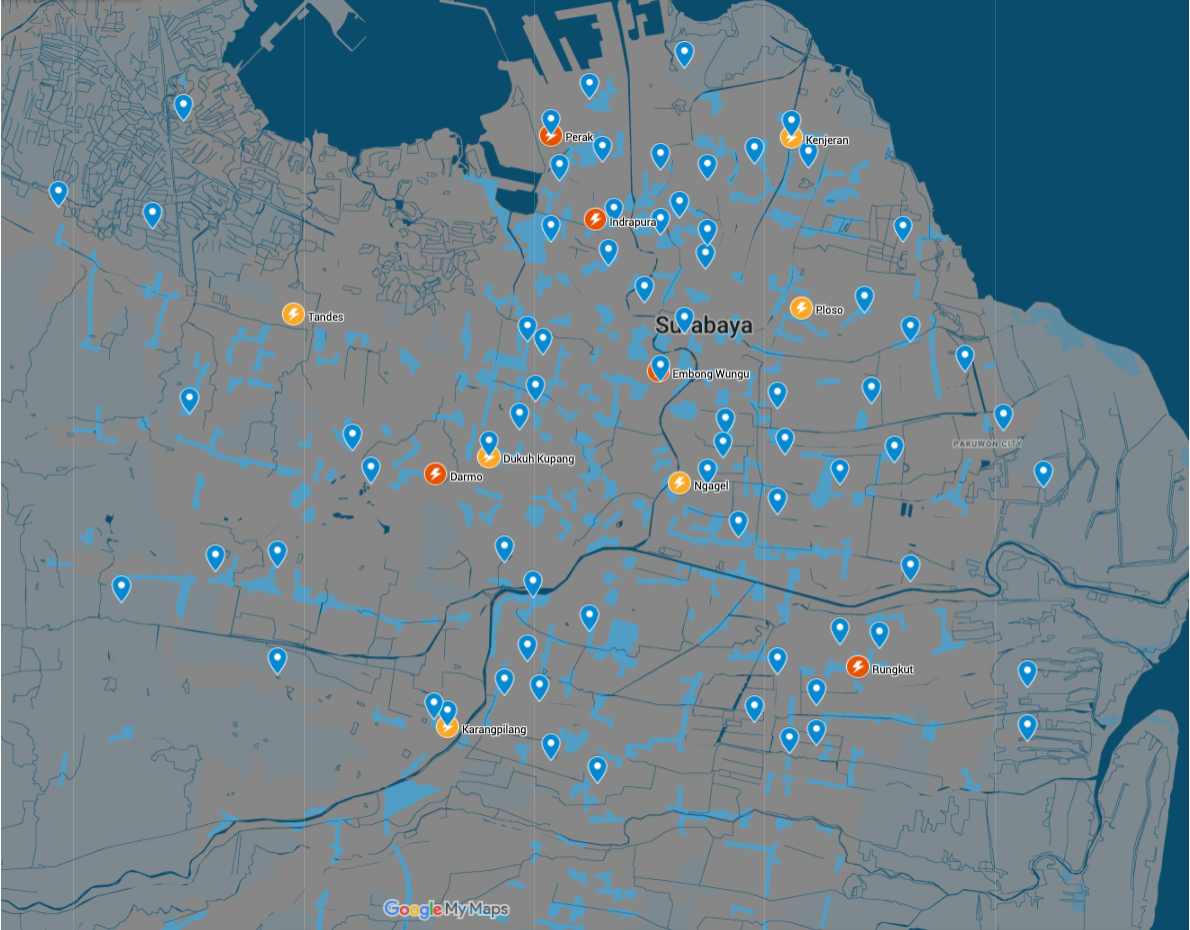}
    \caption{Selected locations for 5 EV charging stations (red markers)}
    \label{fig:sol_5}
\end{figure}

\begin{figure}
    \centering
    \includegraphics[width=\linewidth]{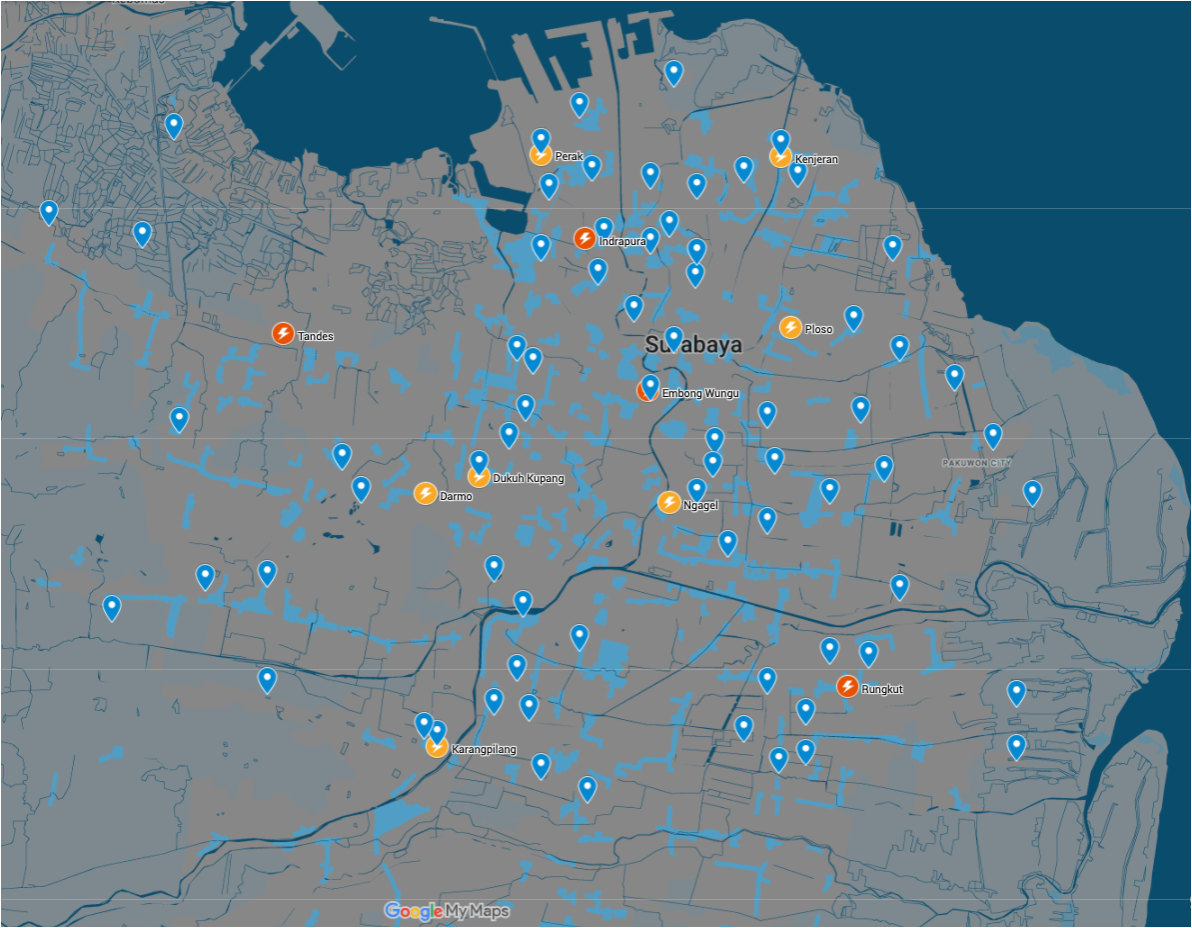}
    \caption{Selected locations for 4 EV charging stations (red markers) }
    \label{fig:sol_4}
\end{figure}

In the experiment, we test multiple values for the time-to-travel threshold $d_{max} = \{25, 30, 35, 40, 45\}$ minutes to assess the sensitivity of the optimal solutions to customer level of service. In this regard, $d_{max}$ parameter (i.e., the maximum distance a customer has to travel to reach an EV charging facility) represents the level of service toward customers. In contrast, lower values mean higher service levels (customers can easily find an EV charging station). In comparison, larger values mean lower service levels (customers must travel further to reach a charging station). Fig. \ref{fig:sumary_opt} shows that a higher service level (lower $d_{max}$ values) requires higher costs, which highlights the tradeoff between service level and total costs. The optimal solution for the baseline problem (without adding a layer of redundancy) is found using OpenSolver \cite{mason2010opensolver} with optimal cost, profit, and revenue reported in Fig. \ref{fig:sumary_opt}. We found that the number of optimal charging stations differ for different $d_{max}$ threshold values (either 4 stations or 5 stations for $d_{max} \in \{25, 30, 35, 40, 45\}$ minutes). Fig. \ref{fig:sol_5} and Fig. \ref{fig:sol_4} show the selected stations as red markers overlaid in Surabaya map for 5-station solution and 5-station solution, respectively.

\begin{figure}[t]
    \centering
    \includegraphics[width=\linewidth]{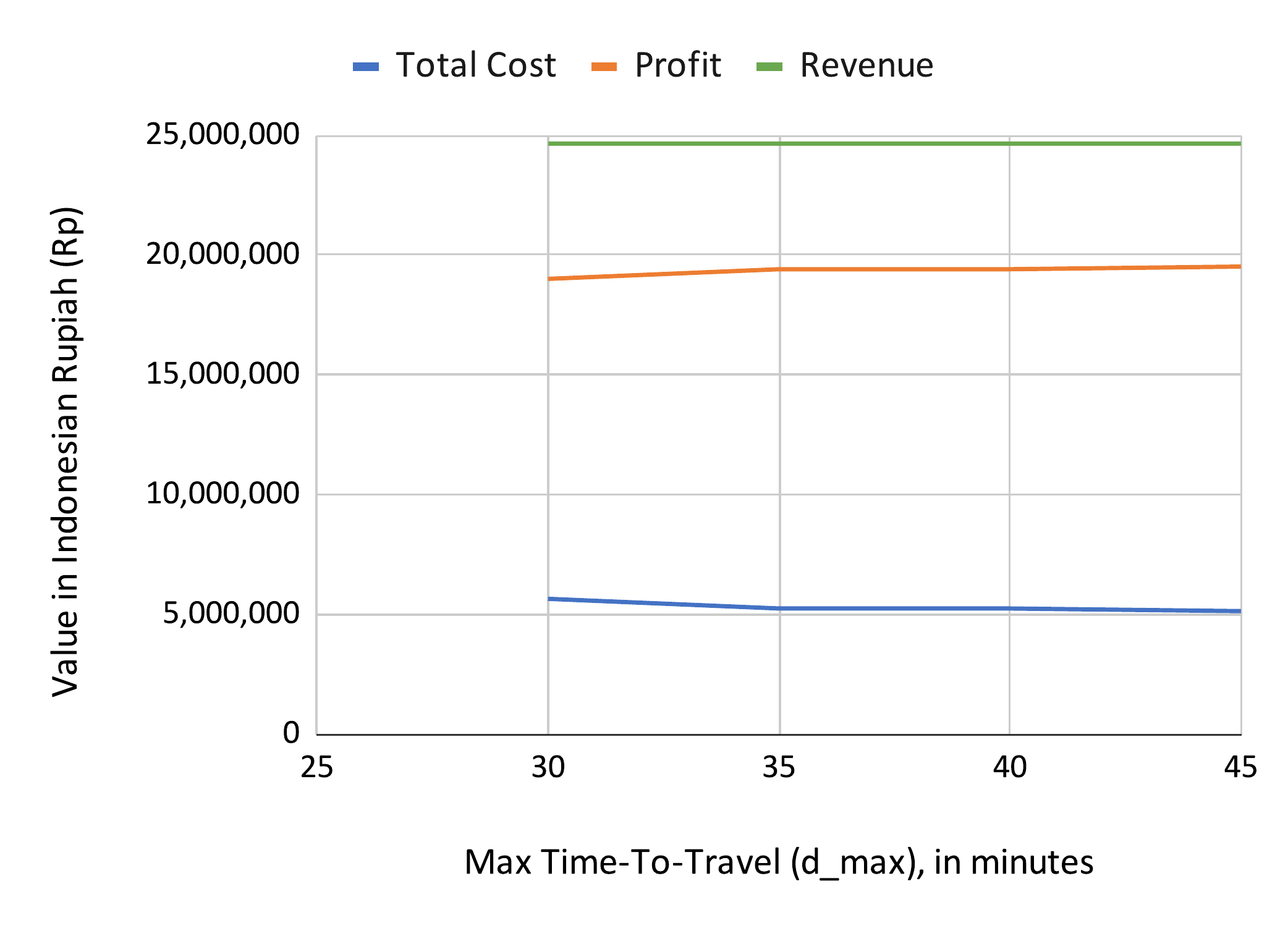}
    \caption{Daily revenue, cost, and profit from the optimal solution of EV charging stations in Surabaya for $d_{max} = \{25, 30, 35, 40, 45\}$ minutes with redundancy ($\geq$2 stations coverage)}
    \label{fig:sumary_opt_2}
\end{figure}

\begin{figure}[t]
    \centering
    \includegraphics[width=\linewidth]{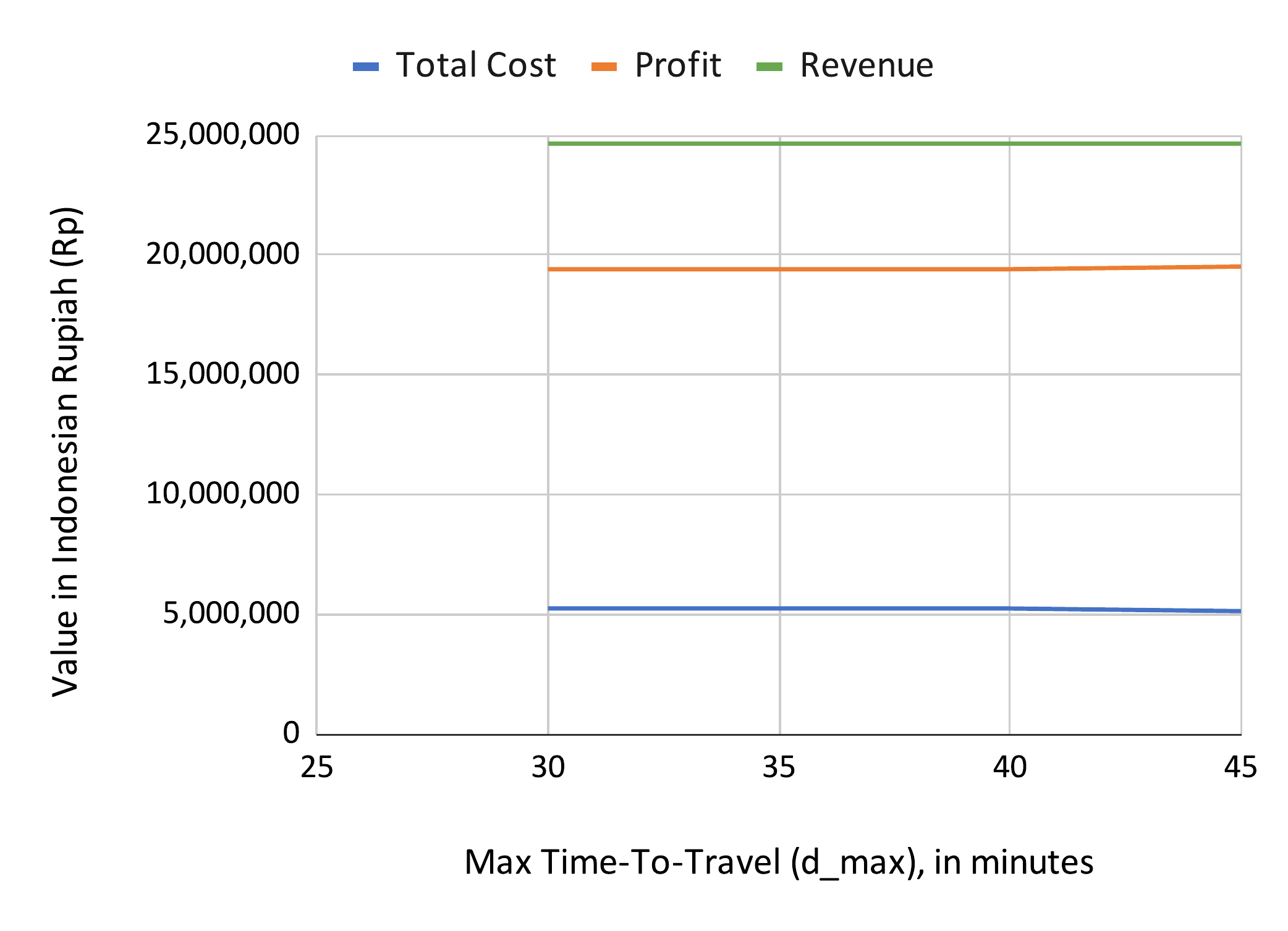}
    \caption{Daily revenue, cost, and profit from the optimal solution of EV charging stations in Surabaya for $d_{max} = \{25, 30, 35, 40, 45\}$ minutes with redundancy ($\geq$3 stations coverage)}
    \label{fig:sumary_opt_3}
\end{figure}

Finally, we study how to increase the overall systems' reliability by adding redundancy layers to buffer against outages and improve customer service levels, incorporating some reliability uncertainties common in developing countries. To account for this, we modify the RHS of constraint (\ref{demandassigned}). Instead of requiring each demand to be covered only by one station, we require 2 or 3 stations to cover each demand. The new revenue, cost, and profit for such a more reliable system is summarized in Fig. \ref{fig:sumary_opt_2} and Fig. \ref{fig:sumary_opt_3}.

\section{Discussions}
\label{sec:discussion}

We first note that we tried to simulate a better level of service situations  ($d_{max}< 25$ minutes) in the experiment, but our model could not find a feasible solution. This is mainly due to high time-to-travel ($d_{ij}$) values due to extensive traffic in Surabaya. Hence, regardless of how the infrastructure is designed, the travel time could not be lowered (unless other forms of traffic intervention are incorporated, such as smart city integration, etc.). Therefore, we only discuss the case for $d_{max} \geq 25$ minutes.

We immediately see from Fig. \ref{fig:sumary_opt} the tradeoff between profit maximization objective with consumer time-to-travel. For example, with a 30-minute threshold for consumer travel time, our model prescribes five charging stations (see Fig. \ref{fig:sol_5} for the locations) to fulfill all demands and a total of 33 units of charging connectors, yielding a revenue of Rp. 24,667,800, a total cost of Rp. 5,654,497, and thus a profit of Rp. 19,013,303. On the other hand, with a more relaxed time-to-travel threshold, say, 35 minutes, our model prescribes only four charging stations (see Fig. \ref{fig:sol_4} for the locations). The total of 33 units of charging stations also, giving the same revenue, but a lower cost of Rp. 5,251,209, resulting in an increased profit Rp. 19,416,591 (2\% higher). We realize that the current improvement in a numerical value is marginal, mainly due to the small projected demand for EVs in Surabaya. In the future, we can scale up our work to cover other larger cities or obtain more accurate EV demand projections, which is expected to rise soon.

We also observe a highly imbalanced demand distribution from the provided data, with more EVs concentrated in the city's wealthier neighborhoods. This encourages sparse solutions, resulting in optimal solutions maximizing the number of charging connectors installed only in one or two of the selected stations and leaving the rest to install only one or two connectors. The different solutions depicted in  Fig. \ref{fig:sol_4} and Fig. \ref{fig:sol_5} highlight how relocating one facility will shift a few other facilities since they are serving a highly density-imbalanced region of the city. This highly imbalanced solution could provide insights for decision-makers to make a more targeted policy to increase the penetration rate of EVs more equally throughout the city. This well-informed policy could potentially exponentiate the positive impacts of EVs for the public, reducing concentrated traffic and pollution and increasing consumer satisfaction overall.

Finally, we highlight the cost of adding layers of redundancy in demand coverage to buffer for service uncertainty, which is particularly important in developing countries since electricity often breaks down, even before adding EVs electricity demand. Fig. \ref{fig:sumary_opt_2} and Fig. \ref{fig:sumary_opt_3} confirm our hypothesis that such redundancy forces the optimizer to output a more reliable network design, hence often comes solutions with higher costs, if such solutions even exist. In our case, there is no such solution for $d_{max}$ = 25 minutes, mainly due to the already constraining travel times $d_{ij}$'s. Thus, we only show the results for $d_{max} \in \{30, 35, 40, 45\}$ in Fig. \ref{fig:sumary_opt_2} and Fig. \ref{fig:sumary_opt_3}. Meanwhile, solutions for $d_{max} > 25$ are at higher costs (7\% increase on average compared to the optimal solution without demand coverage redundancy). With these results, we would advocate using the solution with $d_{max}$ =  25 minutes to achieve a better service level and absorb the 5\% lower profit in the earlier years. If higher profits are demanded, we suggest yielding to the solution with $d_{max} = 35$ minutes at later years. We believe that the government and PLN intend to create a thriving EV ecosystem and consider long-term benefits, thus incorporating demand redundancy in the infrastructure design. We believe the extra costs will pay off as the public adopts EV technologies more widely.

\section{Conclusion and Future Work }
\label{sec:conclusion}

In this study, we present a case concerning EV infrastructure designs for urban areas in developing countries, with data collected from Surabaya, Indonesia. We adopt a maximal covering location problem for our model and solve for the optimal location of the EV charging station and the number of charging connectors installed at each station. Considering 11 alternative locations and 98 sub-districts throughout Surabaya as demand points and projected EV units in each district, our model obtains feasible solutions only when consumers are willing to travel for 25 minutes, given the current traffic conditions in the city. Sensitivity analysis of consumers' time-to-travel reveals that lower-cost solutions are available but force consumers to travel longer. Adding redundancy to the EV infrastructure designs, as an effort to buffer against outages or disruptions, requires higher costs. However, such extra costs might be justifiable in the long term to create a thriving EV ecosystem so that the public can enjoy more environmentally friendly transportation systems and cleaner air. We envision that our model can be applied to incorporate intelligent vehicles as well, optimizing the location of the charging and the microscopic driving behavior that an intelligent vehicle can learn. Such an approach will be one of the subjects of our future studies.



\bibliographystyle{IEEEtran}
\bibliography{IEEEabrv,ref}

\end{document}